# From Principles to Practices: Lessons Learned from Applying Partnership on AI's (PAI) Synthetic Media Framework to 11 Use Cases

Claire R. Leibowicz and Christian H. Cardona[1]
[1]Partnership on AI
San Francisco, CA

## Executive Summary

2023 was the year the world woke up to generative AI, and 2024 is the year policymakers are responding more firmly. In the past year, Taylor Swift fell victim to non-consensual deepfake pornography [3], and a misleading political narrative [22]. A global financial services firm lost $25 million [5] due to a deepfake scam. And politicians around the world [27] have seen their likenesses used to mislead in the lead up to elections. In the U.S., on the heels of a White House Executive Order [23], NIST will be "identifying the existing standards, tools, methods, and practices...for authenticating content and tracking its provenance, [and] labeling synthetic content." Importantly, this policy momentum is taking place alongside real world creation and distribution of synthetic media. Social media platforms, news organizations, dating apps, courts, image generation companies, and more are already navigating a world of AI-generated visuals and sounds, already changing hearts and minds, as policymakers try to catch up. How, then, can AI governance capture the complexity of the synthetic media landscape? How can it attend to synthetic media's myriad uses, ranging from storytelling to privacy preservation, to deception, fraud, and defamation, taking into account the many stakeholders involved in its development, creation, and distribution? And what might it mean to govern synthetic media in a manner that upholds the truth while bolstering freedom of expression? To spur innovation while reducing harm? What follows is the first known collection of diverse examples of the implementation of synthetic media governance that responds to these questions, specifically through [Partnership on AI's (PAI) Responsible Practices for Synthetic Media](#) [13] — a voluntary, normative Framework for creating, distributing, and building technology for synthetic media responsibly, launched in February 2023. In this paper, we present a case bank of real world examples that help operationalize the Framework — highlighting areas synthetic media governance can be applied, augmented, expanded, and refined for use, in practice.

These eleven stakeholders are a seemingly eclectic group; they vary along many axes implicating synthetic media governance. But they're all integral members of a synthetic media ecosystem that requires a blend of technical and humanistic might to benefit society. As Synthesia rightfully notes in their case, "No single stakeholder can enact system-level change without public-private collaboration."

Some of those featured are Builders of technology for synthetic media, while others are Creators, or Distributors. Notably, while civil society organizations are not typically creating, distributing, or building synthetic media (though that's possible [9]), they are included in the case process; they are key actors in the ecosystem of digital media and online information who must have a central role in AI governance development and implementation. Read together, the cases emphasize distinct elements of AI policymaking and seven emergent best practices we explore below. They exemplify key themes that support transparency, safety, expression, and digital dignity online: consent, disclosure, and differentiation between harmful and creative use cases.

| | | |
|---|---|---|
| **Adobe** | Adobe designed its Firefly generative AI model with transparency and disclosure | [Read Adobe's case study](#) |
| **BBC** | BBC used face swapping to anonymize interviewees | [Read BBC R&D's case study](#) |
| **Bumble** | Bumble is preventing malicious AI-generated dating profiles | [Read Bumble's case study](#) |
| **CBC** | CBC News decided against using AI to conceal a news source's identity | [Read CBC Radio-Canada's case study](#) |
| **D-ID** | AI video company D-ID received consent to digitally resurrect victims of domestic violence | [Read D-ID's case study](#) |
| **OpenAI** | OpenAI is building disclosure into every DALL·E image | [Read OpenAI's case study](#) |
| **Respeecher** | Respeecher enables creative uses of its voice-cloning technology while preventing misuse | [Read Respeecher's case study](#) |
| **Synthesia** | AI video startup Synthesia is scaling up content moderation to prevent misuse | [Read Synthesia's case study](#) |
| **TikTok** | TikTok launched new AI labeling policies to prevent misleading content and empower responsible creation | [Read TikTok's case study](#) |
| **WITNESS** | Even the best-intentioned uses of generative AI still need transparency — an analysis by human rights organization WITNESS | [Read WITNESS's case study](#) |
| **PAI** | The risk of synthetic media misuse is growing in global elections — an analysis by PAI | [Read PAI's case study](#) |
| *For a blank version of the template these cases respond to, see [here](#).* | | |

*Table 1.* The list of 11 institutions that submitted case studies reflecting their implementation of PAI's Responsible Practices for Synthetic Media.

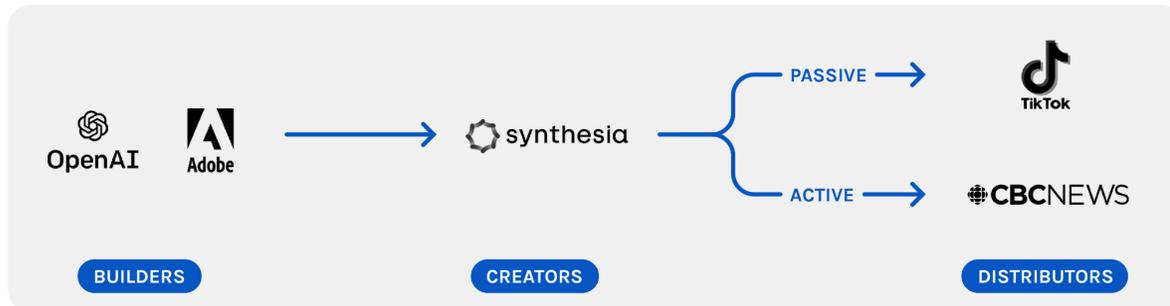

*Figure 1.* An illustrative example of the stakeholder categories depicted in the cases.

| | |
|---|---|
| **Synthetic media** *(also referred to as "generative media")* | Visual, auditory, or multimodal content that has been generated or modified (commonly via artificial intelligence). Such outputs are often highly realistic, would not be identifiable as synthetic to the average person, and may simulate artifacts, persons, or events. |
| **Builders** | Those building technology and infrastructure for synthetic media. |
| **Creators** | Those creating synthetic media. |
| **Distributors** | Those distributing and publishing synthetic media. The Framework identified two types of Distributors: <br> ● <u>Active</u>: Platforms that mostly host first-party content, distribute editorially created synthetic media, and/or report on synthetic media created by others. <br> ● <u>Passive</u>: Platforms that mostly host third-party content. |

*Figure 2.* Definitions of key terms, including synthetic media and the three stakeholder categories differentiating organizations in the cases.

## Theme 1: Creative vs. Malicious Content

Cases that focus on this theme:
*Synthesia*, *Respeecher*, *Bumble*

**Best practice 1: Builders and Creators (not just Distributors) should moderate content to reduce harmful content spreading downstream.**

Several cases respond to a hotly debated question: which stakeholders in the technology pipeline are responsible for content monitoring and moderation?

The debate [7] typically includes some suggesting that moderation by Builders or Creators would stifle innovation and expression, thereby putting too much power in the hands of a few institutions. However, others argue that failing to moderate at the model, technology development, and even infrastructure layer makes it harder to prevent harm downstream.

One of the most public examples of this debate took place far upstream from the institutions featured in this case, yet illustrates these tradeoffs: in 2019, the CEO of Cloudflare, an internet security company, reversed course and terminated 8chan [19], a media platform that allowed "extremists to test out ideas, share violent literature, and cheer on the perpetrators of mass killings." In explaining his decision, and his conflictedness, Cloudflare's CEO mapped out the many institutions undergirding the Internet while questioning [16] how to balance freedom of expression with safety, and the roles they should play in doing so.

Builders, Creators, and policymakers, often face a similar conflict. In our cases, though, several Builder and Creator platforms engaged in normative content moderation (or training data decision making, which in essence affects content development) to support harm mitigation, despite the fact that they are not Distributors of content who are typically those assumed to be responsible [8] for moderating content and for whom much regulatory activity is focused. By doing so, they provide a degree of redundancy in content moderation systems later downstream, possibly minimizing the harmful content that eventually reaches audiences.

For instance, Synthesia, a Builder of synthetic media technology, has implemented detection and moderation capabilities at the point of creation. As they note, "Until recently, most content moderation has happened at the point of distribution: a user of digital creation tools could create content without any restrictions." As with all content moderation, there is inevitably ambiguity in content evaluations, and they differentiate between "obviously harmful content," "obviously harmless content," and "gray zone" content — for which they provide a few examples. However, this moderation taking place before content gets to social media platforms can help support harm mitigation further downstream, though it should be pursued transparently in order to illuminate the often subjective decision-making that takes place when moderating gray area content.

Synthesia describes choices they made about misleading videos about sexual health or cryptocurrency — and how by thwarting their development, they provide meaningful support for eventual social media platform moderation processes that might need to filter out this harmful content. Given such a fast moving field, and the limits of moderation on social media platforms, this might support the actual reduction of harmful content's spread downstream (when done transparently).

Adobe, as a Builder, also took steps to build in technological affordances that would help affect what content is included in and accessible via their models. They are working to enable and protect creators by attaching a "Do Not Train" tag to the metadata of their work so they can ensure that specific content is moderated out of the technology driving synthetic media, and that products further downstream do not then distribute such content.

Notably, while the CBC is not a Builder, their decision as a potential Distributor, in which they chose to not proceed using synthetic media for a storytelling use case, stemmed from the lack of responsibility taken on this task by the software provider — pointing out how Distributors may rely on the content and data decisions made by Builders when thinking about creating and distributing synthetic content.

Just like with more canonical content moderation conducted by Distributors, any moderation should be conducted transparently and mindfully, so as to not stifle innovation and creative expression by those using these tools. We recommend that Builders making content moderation decisions at that stage of development document their actions and disclose their practices, and note [11] that many policies include content moderation transparency stipulations (and they should continue being refined).

**Best practice 2: Balancing creative expression and safety is vital, and means working with content gray areas. Institutions should document decision making about gray area synthetic media cases to drive the field forward, and voluntary commitments alone will not guarantee this documentation is adopted.**

Many cases talked about the need to balance creative expression and safety/harm mitigation. Some provided discrete examples of content that blurs the line between these categories, while others offered only broad acknowledgment of the common tension between these values.

TikTok emphasized their goal of supporting creative expression alongside harm mitigation. WITNESS analyzed the ways in which the creative and harmful might blur, describing how a specific creative project intended to "stir the conscience" could also create unintended harm, describing "a serious possibility that artistic projects that lack prior consent and/or fail to clearly communicate their synthetic nature to audiences [can cause unintentional harm];" Respeecher underscored how they maintain a role as a company devoted to creativity that often serves the entertainment industry and supports accessibility, while also exploring how they acknowledge, and then seek to mitigate, the harmful impacts of synthetic media. Adobe described safety mechanisms in their models that also serve those looking to create using their technology. Even Bumble discussed the often blurry line between those using synthetic media to create fraudulent profiles to defraud users and a non-malicious use like "a member [uploading] a photo of themselves to their profile that has been digitally altered to show them in a location they've never been to before." Synthesia highlights the "gray zone" as part of their analysis, including examples of such content related to sexual health and cryptocurrency contexts.

The cases that explicitly describe gray areas, rather than overarchingly describing them as concepts, help the field understand tradeoffs and how decisions are being made at institutions that implicate the distribution and spread of speech. They have several benefits: they serve as a model for other institutions looking for guidance around exact or analogous scenarios, support broader openness by institutions in this sector, and help users and audiences navigate interactions with the institution in a more informed way.

While it is difficult for institutions to build out a comprehensive set of all of the decisions they have made related to gray area cases, a best practice approach to sharing edge cases and tricky calls must be pursued to ensure that the field is adequately balancing creative expression and harm mitigation. And, of course, different institutions and individuals may have varied perspectives on the appropriate balance between these two considerations. Further, while we encouraged institutions to ground cases in real-world examples of these gray areas, to begin building up these more specific case resources, it will likely take more than this voluntary case study exercise to ensure they are shared at scale.

## Theme 2: Transparency via Disclosure

Cases that focus on this theme:
*[TikTok](), [Adobe](), [BBC](), [OpenAI](), [CBC]()*

**Best practice 3: Builders and Creators should adopt indirect disclosures, or provenance signals, to support Distributors adjudicating content, thereby mitigating harm.**

If Builders implemented more consistent and standardized indirect disclosures [14] — signals for conveying whether a piece of media is AI-generated or AI-modified, based on information about a piece of content's origin that are not user facing — Distributors would have clearer signals that content has been AI-generated, and thus can moderate more easily and support content transparency.

Take, for example, Adobe's exploration of Content Credentials [1], signals that allow consumers of content to understand the origins and changes made to digital files, built off of the C2PA [4] standard, incorporating both invisible watermarking and cryptographically signed metadata. At present, such protocols are baked into Adobe Firefly (and as of this year, OpenAI's DALL·E [12]), thereby enabling social media platforms and content distributors to know when content has been synthesized using those technologies — a step in the right direction for wider adoption.

Baking in such signals of indirect disclosure at the model development stage could also support those distributing content who must deal with identifying harmful synthetic media.

Bumble explicitly describes how such shared standards for indirect disclosure could support them, as a potential Passive Distributor of synthetic media: "[The C2PA standard] would solve detection issues outlined [in the case] and establish trust in the image at every step — all the way from creation to when it's uploaded on a platform. However, this approach would require industry-wide support in order to reliably use it, as well as an invaluable and forward-thinking proof of concept."

TikTok, another Distributor, echoes this sentiment: "If Builders would implement more content provenance/metadata or watermarking techniques in their models, it would greatly benefit our detection and labeling efforts."

While there will always be bad actors ignoring such guidance, and artifact-level signals are only one part of media literacy, these realities should not paralyze the field into passivity; Builders and Creator platforms should adopt indirect disclosures [14] to support Distributors adjudicating content, thereby mitigating harm.

**Best practice 4: Broader public education on synthetic media is required for any of the artifact-level interventions, like labels, to be effective.**

How the field communicates about the impact of methods for evaluating content is just as important as their technical robustness and design.

Several cases explored disclosure methods [14] for supporting audience understanding that content has been AI-generated or not, often through labels attached to individual pieces of content. However, many institutions also highlighted how labels that were applied to specific artifacts did not just have an impact in that particular instance, but were also related to broader societal attitudes and understanding of AI. For instance, societal understanding of what it means to manipulate media, concern that content is synthetic, or belief [21] that labels are applied inaccurately, might affect the impact a specific label attached to a particular artifact has on audiences. This underscores how vital broader literacy and educational campaigns are to field-wide efforts to uphold the truth, and to mitigate harm from synthetic media.

For example, OpenAI recognized that any decisions they made about image provenance signals would exist amidst a context where policymakers and the public might be overconfident in the accuracy and utility of such signals. Furthering public education about the limitations of indirect disclosure methods is a vital prerequisite for their widespread adoption in a manner that serves the public interest. It is also a variable that can affect institutional decision making when implementing different synthetic media governance tactics.

Adobe, writing about their experience designing their Content Credentials, also underscored the ways in which public education is vital for artifact level interventions to work. They highlight the need for future details on "how to accurately create a meaningful and comprehensive disclosure," especially in light of the fact that AI-generated modifications, especially those that are low stakes and do not mislead or cause harm, will soon be so ubiquitous that it could affect how labels, and absence of labels, can signify content credibility or authenticity.

TikTok further emphasized the relationship between artifact-level interventions and broader education, stating, "our disclosure efforts cannot be separated from our efforts to be transparent with our users about what content is created with AI, and to provide users with information and guidance around why we label AIGC [AI-Generated Content], and why we ask them to do the same."

Lack of public education about AI capabilities affected how the CBC, for example, chose to proceed when exploring synthetic media implementation in its reporting for a story that required anonymizing a subject. In other words, partially because audiences were not yet comfortable and well-versed in what synthetic media is and is not, the CBC was understandably reluctant to implement synthetic media in the newsroom; doing so would require broader public literacy before experimenting more readily with AI technologies moving forward. Notably, the BBC did not consider this to be a concern when adopting AI-driven privacy methods for storytelling about Alcoholics Anonymous.

Many in the field agree that we need broader public education, but how society learns about AI, and the impact of such efforts, is rarely described in detail. Based on PAI's previous research [21] on how societal attitudes towards manipulated media labels are often connected to the public's understanding of the institutions involved in their deployment, we are interested in future work that engages with civic institutions — e.g., libraries — and other spaces inhabited by trusted intermediaries that would support audience education about AI. Further, community-centric disclosures that do not get applied solely by technology platforms and large institutions might support greater trust in AI literacy and labels.

Of course, Builders, Creators, and Distributors of synthetic content still have a role to play in educating their audiences about synthetic content and implementing direct disclosures, and they should do so in a way that is open and share access [10] to data about the impact of different disclosure and education approaches.

**Best Practice 5: Creative uses of synthetic media should be labeled, because they might unintentionally cause harm; however, labeling approaches for creative content should be different, and even more mindfully pursued, than those for purely information-rich content.**

Connected to Best Practice 2, one of the major mitigations for ensuring that the line between creative content and harmful content does not blur involves disclosure. Even artistic examples of synthetic media should default to require disclosure — though such disclosures should ultimately preserve, rather than threaten, artistic expression and the creative process.

TikTok and Adobe notably described the development of methods that enable creators to disclose that content has been AI-generated. For TikTok, this is a toggle that creators could use to self-disclose that content has been AI-generated, and in the case of Adobe, it takes shape through Content Credentials. WITNESS's case describes how such disclosure should accompany creative projects developed with synthetic media in order to mitigate the unintended consequences of such content — concepts they've elaborated upon previously [6].

Respeecher's case explained how labeling is useful for creative content but must not come at the expense of creative expression; as they note, for creative contexts like art and entertainment, "overt labeling of a character's voice as synthetic may detract from the user experience, [and] creators have expressed concerns that such labels could disrupt narrative immersion or artistic expression."

The BBC acted as a Creator and Distributor of synthetic media for privacy preservation by obfuscating the faces of subjects in a documentary on Alcoholics Anonymous. They included two different forms of disclosure for that project: in the beginning, the narrator provided auditory disclosure that the project used synthetic media and whenever a subject appeared on the screen, they were accompanied by a caption disclosing that it was an AI-modified image. Other projects have employed privacy preservation via synthetic media, like the documentary film project Welcome to Chechnya [20], which used overhead halos [24] above subjects that were synthetically altered to convey that they had been edited using AI. Creators can therefore consider labeling as part of their creative act.

Respeecher meaningfully highlights the tension that may exist between transparency values and storytelling efforts that require suspension of disbelief. However, taken together, the cases imply a broader benefit to labeling content when, per much of WITNESS's work [6], it is done in a manner that does not detract from the goals of the creative pursuit. Ultimately, creative uses of synthetic media should be labeled in a manner that does not jeopardize the storytelling or artistic goals of the project.

## Theme 3: Consent

Cases that focus on this theme:
[D-ID](), [WITNESS](), [PAI]()

**Best practice 6: Consent for synthetic media should be sought when the likeness of real people is directly involved. And if the subject of synthetic media cannot provide it, Creators still have an obligation to solicit informed consent.**

Consent proved challenging for many institutions across cases. While legal boundaries offer some guidance, responsible creation requires more than achieving the legal bare minimum around topics like intellectual property, and the Framework begins to provide this guidance. WITNESS suggested that consent is even more vital when real people are depicted, advocating for an amendment to the Framework that emphasizes the benefit of "seeking consent when the likeness of real people is directly involved in the input or output of the AI-generation process." They go on to highlight that this should not be mandatory, since there are "some circumstances in which consent may not be pertinent, feasible, or even needed."

The WITNESS case, alongside the D-ID case, dealt with creative projects including real people who could not provide consent — either because they were no longer alive or had been kidnapped — and both provide insight into how to navigate this scenario.

D-ID, writing about a particularly sensitive context — domestic violence — talked to the nuclear family of the featured individual who was no longer alive. Of course, they first needed to deem the social impact goals of educating the public about domestic abuse via the project to be worth the potential emotional tumult of reaching out to families. They even went a step further to bolster consent, allowing the families to actively participate in "co-creating the content and scripts" for the development of the media. This takes informed and active consent — not just about the sheer fact that a creator is using the likeness of their kin, but consent with how that likeness is being used — to the next level.

The WITNESS case also offers guidance for how creators can navigate consent when subjects have been kidnapped or killed. As they note "although there is no clear-cut way to know the preferences of the deceased or missing, contacting relatives, a person's estate, or next-of-kin could be a proactive step in that direction. This approach has been adopted in prior situations, for example by Propuesta Cívica [17], when they constructed a deepfake of murdered journalist Javier Váldez [18]. Interestingly, this example relied upon footage from an archive, presenting interesting questions about the possibility of an archive to grant consent to the creator to use footage of individuals depicted within it, serving as a proxy for the actual individuals' families themselves. Archives of the future might consider stipulations for those submitting material that relate to whether or not the archive can be used for creating synthetic media.

Ultimately, Creators using synthetic media for expressive purposes should seek consent, especially when their projects feature real people, and even if those real people themselves cannot grant consent.

**Best practice 7: When determining how to responsibly receive consent for satirical synthetic content, Creators should consider power dynamics, public vs. private figure status of featured subjects, and the potential for unintended harm from the project.**

As WITNESS has noted in a previous report [25], "for many democratic societies with a tradition of free speech, an individual's public or private status is important when considering whether their consent is necessary before they become the target of a cultural work. Somebody whose words and actions are of legitimate public interest and concern is generally deemed to merit less control over their likeness than an everyday private citizen."

PAI's case study brought this premise into focus, helping provide insight into a thorny question posed in the WITNESS report: in what cases, if any, is consent needed to target individuals in positions of power? The PAI case focused on instances of synthetic media depicting public, political figures around elections — including one that was informational and ostensibly received the figure's consent, and others that did not. While they were not satirical, they did include examples of politicians, individuals who people should be able to deepfake in order to satirize, but not to, as the PAI Framework suggests, "[Manipulate] democratic and political processes, including deceiving a voter into voting for or against a candidate, damaging a candidate's reputation by providing false statements or acts, influencing the outcome of an election via deception, or suppressing voters."

The public status of the politicians in the PAI elections case highlights the ways in which consent might take shape differently depending on the type of political speech one is producing with synthetic media. For example, in the U.S., a jurisdiction with very pronounced speech protections, there are clear categories — like interfering with election processes — that are outright limited. The Biden robocoll example featured in the PAI case is clear because it featured misleading content describing inaccurate processes for voting. It's also possible, though, to imagine a satirist producing a deepfake video of Biden making fun of his gaffes by depicting him in the oval office giving a speech, including content that touches upon topics related to voting practices — thereby presenting a less clear cut scenario than the actual robocall example. This could indeed be satirical, but it could also be used as satirical cover by those looking to mislead the electorate on where to vote. While bad actors might not follow guidance around consent, and in the case of Biden, power and public figure status is very clearcut, those looking to satirize public figures should consider power, status of the individual, and potential harm when determining consent practices. Doing so can support harm mitigation without stifling the project.

The PAI case meaningfully notes, though, that those Building synthetic media like OpenAI have implemented content moderation practices in text-to-image software like DALL·E that prevent individuals from creating synthetic media for public figures, like Barack Obama. This likely derives from OpenAI's risk assessment of the harmful consequences of content creation, but notably stifles creative and satirical expression too. Greater transparency about how OpenAI and others who enact similar filters and content refusals weighed the variables like public vs. private figure status, risk of harm (via threat models), and power against creative potential would support the responsible use of synthetic media. Downstream, Creators must consider these variables when determining consent practices for synthetic content.

## PAI Reflections on the Case Exercise

The case studies in this collection offer the AI field greater transparency into synthetic media governance, highlighting how PAI's Responsible Practices for Synthetic Media [13] can be applied, augmented, expanded, and refined for use in practice. Here, we reflect briefly on several aspects of the governance process, including accountability, transparency, adaptability, and complexity.

**Accountability**
Voluntary frameworks for AI governance are often (understandably) critiqued [15] for providing a facade of rigor and lack of commitment. Many have written [26] on the attempts by technology companies in particular to tout voluntary governance that serves their interests in order to stave off government regulation. This is often true. At the same time, it has become clear through our years of work on synthetic media that in the absence of specific government regulation on synthetic media that can keep pace with the field's development, as well as appetite from stakeholders across sectors for guidance on synthetic media practices that is informed from an ecosystem perspective, PAI could provide a basis for how institutions across the AI field would consider and behave around values like transparency, digital dignity, safety, and expression. This could also provide a foundation featuring policies that have been tested, in practice, that can inform regulatory momentum. Enforcing a reporting requirement was one way for us to remedy the typical lack of accountability for voluntary governance frameworks. We were honest about our inability to strictly mandate guidelines, but we could enforce adherence to providing case studies, where institutions would offer transparency about how they are approaching our guidance. We hoped that doing so might not only deepen adherence to our practices and principles across Framework supporters, but would also help provide transparency about how they did so, thereby providing civil society and the field at large with foundational material; it would also support them holding institutions to account. In the future, we hope to consider how to enable civil society organizations beyond PAI to pressure test and advocate for more specific details from the case writers in media and industry.

**Case Guidance**
These eleven examples provide a rich tapestry of the challenges and opportunities synthetic media governance presents. We were struck by the variety across cases. Some include specific artistic examples, while others focus on broad tradeoffs that implicate AI model development, or specific considerations of news organizations using synthetic media. While they cannot cover the entire surface area of synthetic media impacts, by providing a body of, in essence, case law for synthetic media, we offer the field a starting point for navigating their own synthetic media challenges. For example, if one is navigating a creative project that deals with posthumous consent, they can consult the D-ID or WITNESS cases. Notably, these cases required enormous effort and time across PAI staff and Framework supporters, and we are interested in developing methods for collecting cases and instances of synthetic media decision making that might not require long-form writing — something akin to the AI Incident Database [2] that was created at PAI. Starting with the level of depth exemplified in the cases, though, provides a useful foundation for understanding the complexity of case examples featuring synthetic media challenges and opportunities, and also allows us to put them in context and dialogue with other actors in the synthetic media pipeline.

**Framework Adaptability and Refinement**
Another benefit of the case process was pointing out ways that the Framework can be augmented or adapted over time. A key principle of the Framework's launch was that, in direct response to the rapid pace of AI development, we would revise the Framework. Several details emerged throughout the case process that will inform future versions of the Framework, including but not limited to:

- Proposing that Builders, Creators, and Distributors should enable and/or use more than one disclosure mechanism to offset shortcomings.
- Including a provision to highlight the need to develop standardized and interoperable solutions for disclosure.
- Suggesting clear guidance on how to label different creative types of synthetic media.
- Offering details on consent that can help address gray-area cases when dealing with the likeness of a deceased or missing person.
- Providing insight into seeking consent from real people whose images are included in media.
- Describing clearer thresholds for what makes something "synthetic enough" to be directly disclosed.

**Institutional Transparency**

The transparency afforded by these cases is a step in the right direction for the field — of course, the cases reveal instances of synthetic media development, creation, and distribution that shed light on institutional practices and tactics. In addition, the manner in which the institutions described and analyzed their decision making, and chose to share it, also offers transparency into institutional practices. One of the benefits, and challenges, from an open-ended case template was that institutions had quite a bit of flexibility in how they could focus, and describe their cases; one could focus on something as broad as general policy development or as specific as a particular gray area case that prompted debate, with varying levels of detail (though PAI pushed emphatically for more detail, across cases, with methods we will describe in more depth in future reporting). This flexibility was both practical (to enable us to learn more about how institutions would respond to our first foray into case studies of this sort) and useful (since we were interested in learning more about many levels of implementation of Framework principles and practices).

We were particularly heartened by the cases that offered frank introspection and were written in a manner that acknowledged when they changed course, and meaningfully, described why — like in the case of OpenAI describing how they navigated their text detection decision making. This is the type of honest reflection we hope to promote, stylistically and substantively, in all future versions of the cases.

**Complexity**

One of the trickiest realities of the case study effort is the extent to which universal themes emerged, but so too did very unique, specific elements come through for each case. Ecosystem actors face similar value tradeoffs regardless of their positions in the synthetic media pipeline, but their specific institutional considerations — and even specific case considerations — need to guide their responses to those tradeoffs. This makes the job of those creating frameworks that move beyond merely stating "do no harm" and thus apply across specific cases and sectors quite tricky, as they need to balance degrees of flexibility and specificity that prove useful to the real world examples the field encounters. Our hope is that this exercise meaningfully highlights the complexities of synthetic media governance, while also producing tangible recommendations that work across cases and underscore the utility of an ecosystem approach. While these cases are focused on synthetic media, they touch vast societal dynamics ranging from freedom of speech, the meaning of harm, transparency, creative endeavor, and consent — topics that each warrant their own specific analyses. The utility of a case exercise, then, is not only the coherent themes across cases, but also the distinct facets that take shape in individual cases. Thus, we encourage institutions to pay attention to their distinct considerations when making decisions about synthetic media governance. Meaningful synthetic media governance should be useful for specific institutions, as well as broader stakeholders.

**Where We Go From Here**
Government regulation and policy are key complements to the Synthetic Media Framework and governance activities at PAI more broadly. Our hope is that policymakers not only learn from the emergent best practices in these cases, but also consider:

- The interconnectedness of Builders, Creators, and Distributors in the synthetic media pipeline.
- The need for flexibility, and specificity, in synthetic media policymaking.
- How the narrative considerations accompanying policymaking focused on synthetic media transparency may impact their efficacy — for example, how is the impact of something like indirect disclosure's adoption conveyed to the public?
- The need for synthetic media policy to adapt over time.
- The ways in which different sectors — social media platforms, media institutions, dating applications, synthetic media creator platforms, AI technology companies — might require distinct recommendations for how to enact certain values.
- The centrality of consent, transparency, support for creative expression, and harm mitigation to synthetic media policymaking.

We plan to report in more depth on PAI's analysis of the case study process soon. In the coming months, PAI will be working to analyze and refine this case study process for the eight additional institutions who have joined the Framework. Through our engagement with policymakers, including the NIST Safety Institute in the US, we will be sharing insights from these case studies and this exercise in synthetic media governance with the policy community. And further, we hope to drill deeper into some of the open questions underscored in the cases — just as we further operationalized key elements of the Framework, like indirect disclosure methods [17], through multistakeholder convening and collaboration.

We look forward to sharing more insights about the cases, how they were developed, and how they have impacted the field in the coming months.

# Acknowledgements

We thank Rebecca Finlay and Stephanie Bell for their comments on earlier versions of this draft, as well as all of the organizations that wrote cases as part of PAI's Synthetic Media Framework.

[25] WITNESS. 2023. Just Joking: Deepfakes, Satire, and the Politics of Synthetic Media. https://cocreationstudio.mit.edu/just-joking/

[26] Alyssa Wong. 2023. Regulatory gaps and democratic oversight: On AI and self regulation. https://srinstitute.utoronto.ca/news/tech-self-regulation-democratic-oversight

[27] Pranshu Verma, Meryl Kornfield. 2024. Democratic operative admits to commissioning Biden AI robocall in New Hampshire. https://www.washingtonpost.com/technology/2024/02/26/ai-robocall-biden-new-hampshire/and-use-of-artificial-intelligence/